\documentclass{iopjournal}

\usepackage{graphicx}
\usepackage{dcolumn}
\usepackage{bm}
\usepackage{subcaption}
\usepackage[utf8]{inputenc}
\usepackage[T1]{fontenc}
\usepackage{mathptmx}
\usepackage{newunicodechar}
\usepackage{booktabs}
\usepackage{float}
\newunicodechar{₂}{$_2$}
\usepackage{etoolbox}
\usepackage{xcolor}
\usepackage[normalem]{ulem}
\usepackage{siunitx}
\usepackage{amsmath}
\usepackage{amssymb}

\usepackage[
  backend=biber,
  style=phys,
  articletitle=false, 
  sorting=none,
  sortcites=true,
]{biblatex}
\begin{document}

\articletype{Paper} 

\title{Ferroelastic instability in rutile TiO$_2$ and thermodynamic suppression of the CaCl$_2$-type phase}

\author{Jared Pohlmann$^1$\orcid{0009-0005-7084-9506}, Anjy-Joe Olatunbosun$^1$\orcid{0009-0008-3134-956X} and Kenneth Park$^{1,*}$\orcid{0000-0002-3675-1776}}

\affil{$^1$Department of Physics and Astronomy, Baylor University, Waco, Texas, USA}


\affil{$^*$Author to whom any correspondence should be addressed.}

\email{kenneth\_park@baylor.edu}


\begin{abstract}
We investigate the role of the CaCl$_2$-type ($Pnnm$) phase in the high-pressure transformation of rutile TiO$_2$, whose experimental signature has remained elusive. While analogies with other rutile-type oxides suggest such an intermediate, TiO$_2$ typically exhibits a direct transformation to higher-coordination phases such as baddeleyite. Using an all-electron density functional framework combined with density functional perturbation theory, we show that rutile TiO$_2$ undergoes a ferroelastic instability characterized by the development of an orthorhombic strain and a double-well energy landscape at $\sim$13.5~GPa. This instability is associated with the softening of the $C_{11} - C_{12}$ elastic combination and the condensation of a $B_{1g}$ phonon mode, involving coordinated rotations of TiO$_6$ octahedra that lower the symmetry to the $Pnnm$ structure. Despite this clear elastic and dynamical pathway, enthalpy calculations show that the CaCl$_2$-type phase is only weakly stabilized relative to rutile and remains energetically unfavorable compared to competing columbite and baddeleyite phases. Consequently, the $Pnnm$ phase does not emerge as a stable high-pressure polymorph but instead exists as a transient or weakly metastable intermediate. These results demonstrate that the CaCl$_2$-type phase represents the intrinsic ferroelastic response of rutile TiO$_2$, yet is suppressed by thermodynamic competition, providing a consistent and unified explanation for its elusive experimental observation.
\end{abstract}

\section{Introduction}

Titanium dioxide (TiO$_2$) is widely utilized in industrial and technological applications owing to its attractive optical and electronic properties combined with structural robustness \parencite{stephen_lourduraj_a0cd34f6, anca_diana_racovita_78b88d12, ragavi_priyadharshani_raja_05920473}. Its high refractive index makes TiO$_2$ a common white pigment, while its photocatalytic activity under UV irradiation enables applications in environmental remediation and self-cleaning surfaces \parencite{eneyew_tilahun_bekele_59d5156e, zohaib_razzaq_36510228, nyiko_m__chauke_3e9d3da6}. In addition, TiO$_2$ serves as an electron-transport material in solar cells and finds use in sensors, UV blockers, and advanced ceramics.

Under compression, TiO$_2$ exhibits rich polymorphism, generally following a sequence of increasing titanium coordination. The ambient-pressure rutile structure ($P4_2/mnm$) remains stable up to $\sim$12~GPa, where transformation to the seven-coordinated monoclinic baddeleyite phase ($P2_1/c$) begins \parencite{hiromasa_sato_75db7ff0}. Upon further compression (19–36~GPa), the system enters the orthorhombic OI phase ($Pbca$) \parencite{natalia_dubrovinskaia_817d86ea}, and at higher pressures the eight-coordinated cotunnite phase ($Pnma$) becomes stable \parencite{dubrovinsky2001hardest, nishio2010stability}. On decompression, strong hysteresis is observed: rather than reverting directly to rutile, the material typically transforms into the orthorhombic columbite ($\alpha$-PbO$_2$-type, $Pbcn$) phase, which is often recovered metastably due to kinetic barriers \parencite{l__gerward_0d36ab58, a__g__christy_b8df758d, liu1979high}.

Early insights into these transformations were obtained from shock-wave experiments, which identified major structural changes near $\sim$30~GPa accompanied by significant volume collapse \parencite{r__g__mcqueen_f45d41f2}. Subsequent static compression studies in diamond anvil cells refined these observations, revealing that transformations can initiate at pressures as low as $\sim$15~GPa and exhibit strong anisotropy depending on crystallographic direction \parencite{ronald_k__linde_74bbd3c2, yasuhiko_syono_6d92e51f}. These results suggested the reconstructive nature of the transformation to higher coordination, although the detailed mechanisms and intermediate structures remained uncertain.

Rutile-type dioxides more broadly display similar pressure-induced transformations toward higher coordination. A common feature among these materials is the emergence of intermediate structures, including the columbite ($\alpha$-PbO$_2$-type) phase and, in many cases, an orthorhombic CaCl$_2$-type ($Pnnm$) phase. For example, MnO$_2$, CrO$_2$, and SnO$_2$ exhibit ferroelastic transitions to the CaCl$_2$-type structure at modest pressures, often accompanied by elastic softening and symmetry lowering \parencite{haines1995second, b__r__maddox_82081e3d, julien_haines_531bbe53}. Similar behavior is observed in systems such as GeO$_2$ and stishovite SiO$_2$, where the CaCl$_2$-type phase serves as an intermediate linking rutile to denser polymorphs \parencite{haines2000structural, d__andrault_5b56bfdb}. Structurally, this phase involves coordinated rotations of MO$_6$ octahedra and is widely regarded as the intrinsic ferroelastic instability of the rutile lattice.

In TiO$_2$, however, evidence for such a CaCl$_2$-type intermediate remains indirect. Raman spectroscopy revealed a softening of the $B_{1g}$ mode under pressure \parencite{nicol1971raman}, which was later identified as a precursor to a rutile-to-CaCl$_2$-type instability \parencite{nagel1971pressure}. Despite this, direct structural identification has proven difficult, and experiments often observe a direct transformation to higher-coordination phases. Recent simulations have suggested a two-step pathway involving a transient CaCl$_2$-type configuration \parencite{yu_liu_a13516ce}, but the small magnitude of the associated distortions makes experimental detection challenging. Consequently, the CaCl$_2$-type phase has been described as a ``ghostly'' intermediate—predicted theoretically but rarely observed.

Early first-principles work by Montanari and Harrison \parencite{montanari2004pressure} identified a pressure-induced softening of the $B_{1g}$ mode and a collapse of the shear modulus $C_{11} - C_{12}$, suggesting a possible instability toward the CaCl$_2$-type ($Pnnm$) structure near $\sim$13~GPa. However, they emphasized that no experimental evidence for this phase existed and that the transition could not be conclusively established. In particular, the proximity of competing transformations and the lack of clear structural identification left the role of the CaCl$_2$-type phase unresolved.

These considerations indicate that, despite growing theoretical support, the thermodynamic stability and physical role of the $Pnnm$ phase in TiO$_2$ remain unclear. It is therefore not known whether this phase represents an equilibrium intermediate under hydrostatic conditions or a transient distortion that is bypassed in favor of more stable polymorphs. This motivates a systematic first-principles investigation.

In this work, we employ density functional theory (DFT) to study rutile TiO$_2$ under hydrostatic pressure. We first evaluate its mechanical stability with respect to orthorhombic distortions, followed by an analysis of vibrational properties to identify soft phonon modes associated with structural instability. Finally, we compare the enthalpies of competing phases up to $\sim$30~GPa to clarify their relative stability and to elucidate the origin of the elusive CaCl$_2$-type intermediate in TiO$_2$.

\section{Methodology}
\subsection{Electronic structure calculations}

DFT calculations were performed using the all-electron, full-potential augmented plane wave plus local orbitals (APW+lo) method as implemented in the \textsc{WIEN2k} package \parencite{peter_blaha_76330271}. In this framework, the unit cell is partitioned into non-overlapping muffin-tin (MT) spheres centered on atomic sites and an interstitial region. Inside the MT spheres, Kohn–Sham wavefunctions are expanded in spherical harmonics multiplied by radial functions, while plane waves are used in the interstitial region.

The basis set was constructed using the linearized APW (LAPW) scheme, in which radial functions and their energy derivatives are combined to form a variational basis \parencite{singh2006planewaves}. To improve the variational flexibility, particularly for semicore and high-lying states, additional local orbitals were included, yielding the APW+lo formalism. These local orbitals are confined within the MT spheres and enhance the description of states not well captured by the standard LAPW basis.

Core and valence states were separated at an energy cutoff of $-6.0$~Ry. For Ti, the $3s$ and $3p$ states were treated as semicore, while Ti $3d$, $4s$ and O $2s$, $2p$ states were treated as valence. Core states were treated fully relativistically, while valence states were described within the scalar-relativistic approximation. Exchange--correlation effects were treated using the Perdew--Burke--Ernzerhof (PBE) functional within the generalized gradient approximation (GGA) \parencite{john_p__perdew_9e66d033}.

The muffin-tin radii were chosen as $R_{\mathrm{MT}} = 1.78\,a_B$ for Ti and $1.61\,a_B$ for O. The basis-set size was controlled by $R_{\mathrm{MT}}K_{\mathrm{max}} = 9$, ensuring convergence of relative total energies between competing phases. The Fourier expansion of the charge density in the interstitial region was truncated at $G_{\mathrm{max}} = 20$~Bohr$^{-1}$, and lattice harmonics inside the MT spheres were included up to $L_{\mathrm{max}} = 10$.

Self-consistent calculations were converged to $10^{-5}$~Ry in total energy and $10^{-3}$~e in charge density. Structural relaxations were performed until Hellmann–Feynman forces were smaller than $0.5$~mRy/Bohr. These parameters provide sufficient numerical accuracy to resolve the small energy differences relevant for phase stability in TiO$_2$, consistent with our previous work \parencite{pohlmann2025stability}.

\subsection{Elastic properties}

Elastic properties were calculated using the \textsc{IRELAST} package \parencite{m__jamal_baf17896} interfaced with \textsc{WIEN2k}. Strained lattice vectors $\mathbf{R}'$ were obtained from the equilibrium lattice $\mathbf{R}$ through a deformation matrix $\mathbf{D}$, where the strained lattice vectors are obtained as $\mathbf{R'} = \mathbf{R} \cdot \mathbf{D}$ with $\mathbf{D} = \mathbf{I} + \mathbf{\varepsilon}$, and $\mathbf{\varepsilon}$ denotes the infinitesimal strain tensor.

The total energy under strain was expanded to second order as
\begin{equation}
E(V,\boldsymbol{\varepsilon}) = E(V_0) + V_0 \sum_{i,j} \sigma_{ij} \varepsilon_{ij} + \frac{V_0}{2} \sum_{i,j,k,l} C_{ijkl} \varepsilon_{ij} \varepsilon_{kl},
\end{equation}
where $V_0$ is the equilibrium volume, $\sigma_{ij}$ is the stress tensor, and $C_{ijkl}$ are the second-order elastic (stiffness) constants \parencite{wallace1972thermodynamics}.

In practice, the tensors were expressed in Voigt notation, where the indices are mapped as $xx \rightarrow 1$, $yy \rightarrow 2$, $zz \rightarrow 3$, $yz$ ($zy$) $\rightarrow 4$, $xz$ ($zx$) $\rightarrow 5$, and $xy$ ($yx$) $\rightarrow 6$. In this representation, the strain, stress, and stiffness tensors are written as $\varepsilon_i$, $\sigma_i$, and $C_{ij}$ with $i,j = 1,\ldots,6$.

\subsection{Phonon calculations}

Vibrational properties were calculated using density functional perturbation theory (DFPT) as implemented in the \textsc{Quantum ESPRESSO} package \parencite{paolo_giannozzi_2489de97}. In DFPT, phonon frequencies are obtained from the dynamical matrix, defined as the second derivative of the total energy with respect to atomic displacements and evaluated at wavevector $\mathbf{q}$ \parencite{stefano_baroni_ae84e183}. Imaginary frequencies indicate dynamical instabilities.

The DFPT calculations were performed within the local density approximation (LDA). This choice is motivated by known inconsistencies in the stability of rutile TiO$_2$ within PBE-GGA, which depend sensitively on the treatment of Ti semicore states in pseudopotentials \parencite{keith_refson_d669bbf7, krishna_k_ghose_79f1c4dd}. LDA provides a stable description of the rutile phase and yields vibrational frequencies in good agreement with experiment.

Ultrasoft pseudopotentials were employed, with a plane-wave cutoff of 90~Ry and a charge density cutoff of 720~Ry. Brillouin zone integrations were performed using a Monkhorst--Pack $k$-point mesh of $8 \times 8 \times 8$, ensuring convergence of total energies and phonon frequencies. Electronic self-consistency was achieved with a convergence threshold of $10^{-12}$~Ry, and the phonon convergence threshold was set to $10^{-14}$. Dynamical matrices were computed on an $8 \times 8 \times 8$ $\mathbf{q}$-point mesh to ensure accurate interpolation of phonon dispersions.

Long-range electrostatic interactions were treated by explicitly computing the Born effective charge tensors and high-frequency dielectric tensor, enabling proper description of longitudinal-optical (LO) and transverse-optical (TO) mode splitting near the $\Gamma$ point.

\section{Results \& Discussion}
For rutile TiO$_2$, the tetragonal symmetry gives rise to six independent elastic constants: $C_{11}, C_{33}, C_{44}, C_{66}, C_{12}$, and $C_{13}$. The values calculated from the  zero-pressure structure (Supplementary Materials) are listed in Table~\ref{table: elastic_constants}, together with representative experimental data. Among these, $C_{33}$ is by far the largest, nearly twice the magnitude of the next largest component, $C_{11}$ \parencite{murli_h__manghnani_8a9c7819, m__grimsditch_1a2699a6, donald_g__isaak_5443c786}. This pronounced anisotropy reflects the structural characteristics of rutile, where the $c$-axis aligns with chains of edge-sharing TiO$_6$ octahedra, resulting in significantly enhanced resistance to deformation along this direction \parencite{baur2007rutile}.

Overall, our calculated elastic constants are in good agreement with experimental values. In particular, $C_{33}$ and $C_{13}$ show excellent agreement, while $C_{66}$ is overestimated by approximately 10--12\%. The remaining constants are underestimated by about 3--6\%. This systematic underestimation is consistent with the known tendency of the PBE functional to slightly overestimate equilibrium lattice parameters, leading to a corresponding softening of elastic responses. The present results are also consistent with previous DFT studies using GGA exchange-correlation functional \parencite{yao2007ab, ma2009pressure, wu2010structural, qi_jun_liu_83e53746}. 

\begin{figure}[htbp]
\centering
\includegraphics[width=0.8\textwidth]{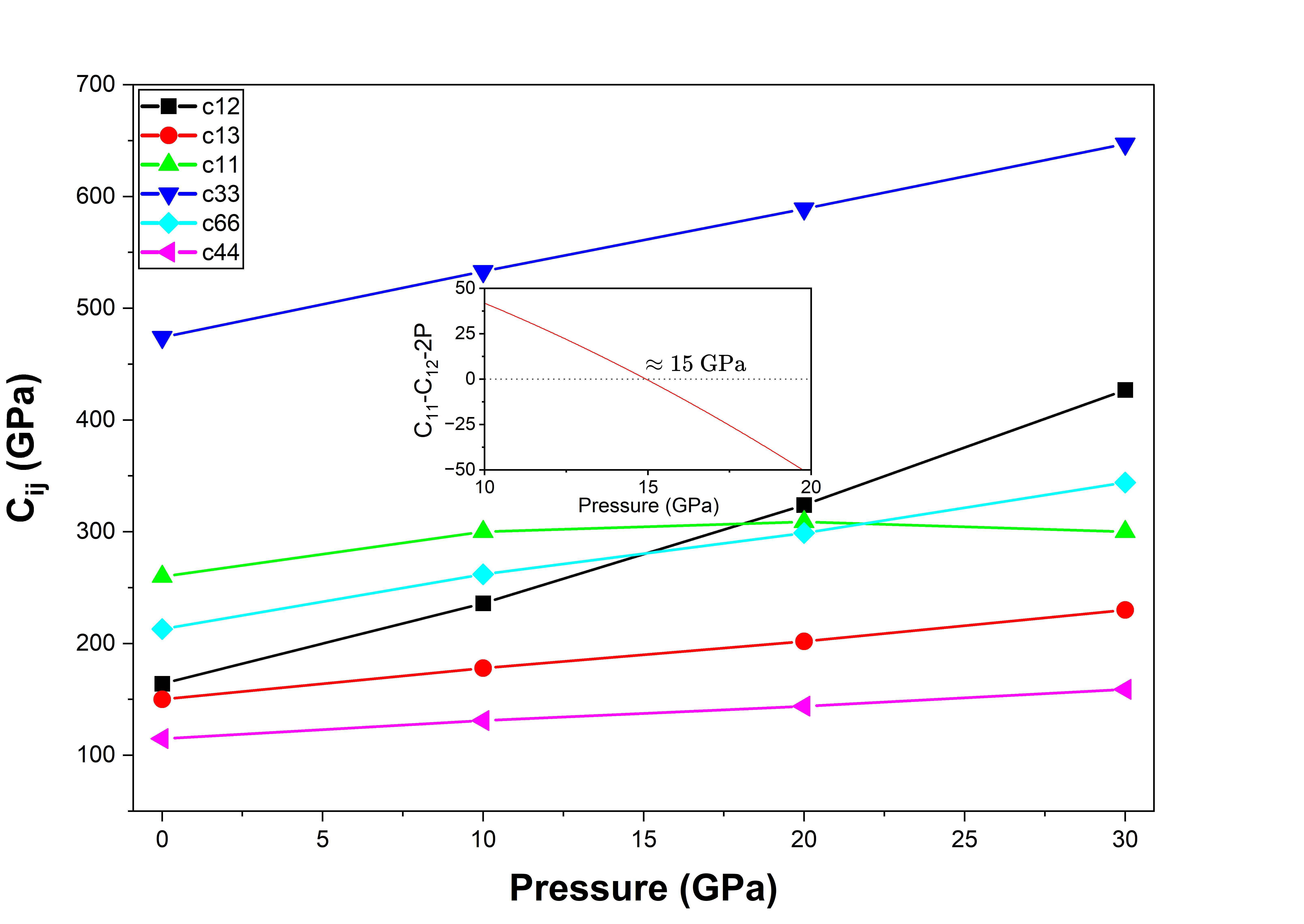}
\caption{Pressure-dependent elastic constants $C_{ij}$. The inset shows the interpolated pressure at which $C_{11} - C_{12} -2P = 0$.}
\label{fig:cij_pressure}
\end{figure}

\begin{table}[H]
\caption{Calculated elastic constants compared with experimental ranges for the tetragonal system. All values are in GPa.}
\label{table: elastic_constants}
\centering
\renewcommand{\arraystretch}{1.4}
\setlength{\tabcolsep}{10pt}

\begin{tabular}{@{}c S[table-format=3.0] l@{}}
\toprule
\textbf{$C_{ij}$} & {\textbf{This Work PBE}} & \textbf{Experiment} \\
\midrule
$C_{11}$ & 260 & 270\textsuperscript{a}, 267\textsuperscript{b}, 268\textsuperscript{c} \\
$C_{12}$ & 164 & 176\textsuperscript{a}, 181\textsuperscript{b}, 175\textsuperscript{c} \\
$C_{13}$ & 150 & 148\textsuperscript{a}, 147\textsuperscript{b,c} \\
$C_{33}$ & 474 & 484\textsuperscript{a,c}, 479\textsuperscript{b} \\
$C_{44}$ & 115 & 124\textsuperscript{a,c}, 123\textsuperscript{b} \\
$C_{66}$ & 213 & 193\textsuperscript{a}, 189\textsuperscript{b}, 190\textsuperscript{c} \\
\bottomrule
\end{tabular}

\noindent{\footnotesize{$^{a}$ Ref. \cite{murli_h__manghnani_8a9c7819}; 
$^{b}$ Ref. \cite{m__grimsditch_1a2699a6}; 
$^{c}$ Ref. \cite{donald_g__isaak_5443c786}}}
\end{table}

Figure~\ref{fig:cij_pressure} shows the evolution of the elastic constants under increasing hydrostatic pressure. Most components increase monotonically, indicating the expected stiffening of the lattice under compression. In contrast, $C_{11}$ (green triangles) exhibits a qualitatively different behavior. After a slight initial increase, it saturates and subsequently decreases beyond $\sim$10--15~GPa. This behavior is more naturally interpreted in terms of the combination $C_{11} - C_{12}$, which directly governs the tetragonal shear stability.

Mechanical stability requires the elastic stiffness tensor to remain positive definite. At zero pressure, this condition leads to the conventional Born stability criteria. Under finite hydrostatic pressure $P$, these conditions must be modified to account for the external stress, leading to generalized stability criteria \parencite{g_v__sin_ko_4744d961}. For the tetragonal Laue class I ($4/mmm$, $422$, $\bar{4}2m$, $4mm$), the relevant conditions are

\begin{equation}
\begin{split}
C_{11} - C_{12} - 2P > 0, \quad C_{44} - P > 0, \quad C_{66} - P > 0, \\
\qquad C_{11} + C_{33} - 2C_{13} - 4P > 0, \\
\qquad 2C_{11} + 2C_{12} + C_{33} + 4C_{13} + 3P > 0.
\end{split}
\end{equation}

As shown in Fig.~\ref{fig:cij_pressure}, the criterion $C_{11} - C_{12} - 2P > 0$ is the first to be violated with increasing pressure. Extrapolation of the calculated data yields a critical pressure of $\sim 15$~GPa (see inset). This value is significantly lower than the $\sim$19.2~GPa reported by Liu \textit{et al.} \parencite{qi_jun_liu_83e53746}, although the overall pressure dependence of the elastic constants is qualitatively consistent between the two studies.

The discrepancy can be traced primarily to differences in the calculated values of $C_{11}$ at intermediate pressures. In Ref.~\parencite{qi_jun_liu_83e53746}, $C_{11}$ is substantially larger at 10 and 20~GPa (325 and 350~GPa, respectively) compared to the present results (305 and 304~GPa). Conversely, their $C_{11}$ becomes significantly smaller at higher pressure (30~GPa), where it drops to $\sim$190~GPa, compared to our value of 275~GPa. These differences highlight the sensitivity of the mechanical stability threshold to the detailed pressure evolution of $C_{11}$.

The quantitative discrepancy likely arises from methodological differences between the two studies. In the present work, elastic constants are obtained from an all-electron APW+lo framework, whereas Liu \textit{et al.} \parencite{qi_jun_liu_83e53746} employed a plane-wave pseudopotential approach. Although semicore Ti states were included in their valence configuration, residual pseudopotential approximations may influence the stress response, particularly under compression. In addition, their plane-wave cutoff (380 eV) is relatively modest for accurately converging elastic properties of transition-metal oxides, which are known to be sensitive to basis-set completeness. Differences in exchange–correlation functional (PW91 vs PBE) may also contribute, although these effects are expected to be comparatively minor relative to basis-set and pseudopotential convergence. Taken together, these factors provide a consistent explanation for the observed differences in the pressure evolution of $C_{11}$ and the resulting stability threshold.

\begin{figure}[htbp]
\centering
\begin{subfigure}[b]{0.8\textwidth}
    \centering
    \includegraphics[width=0.8\textwidth]{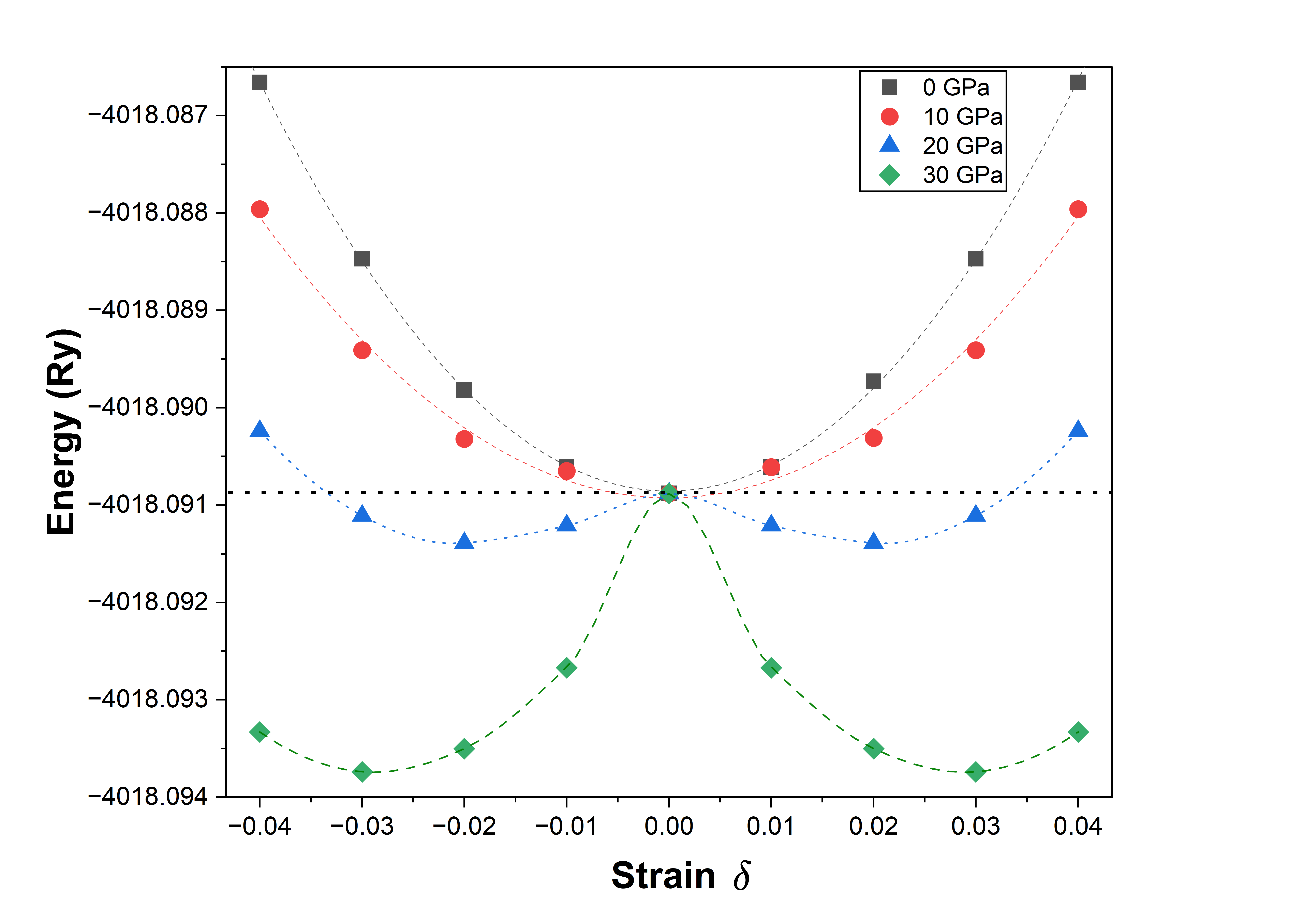} 
    \caption{}
    \label{fig:strain_energy}
\end{subfigure}
\begin{subfigure}[b]{0.8\textwidth}
    \centering
    \includegraphics[width=0.8\textwidth]{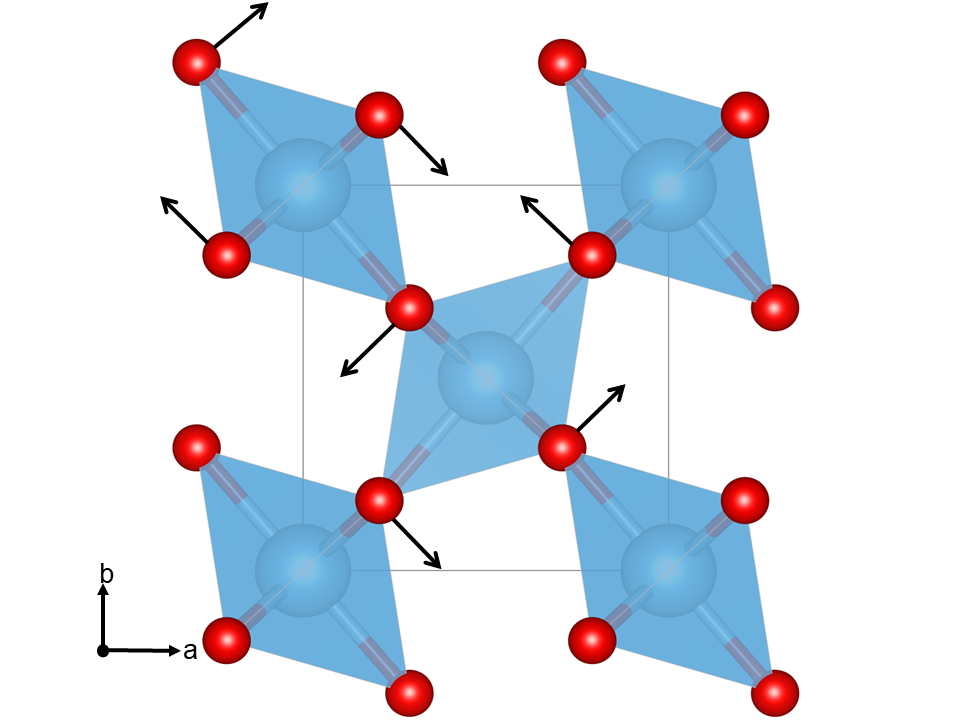}
    \caption{}
    \label{fig:tio2_pnnm}
\end{subfigure}
\caption{(a) Energy as a function of the orthorhombic strain $\delta$ at selected pressure values. For clarity, the energy for $10$, $20$, and $30$ GPa is shifted down by $0.005$, $0.021$, and $0.041$ Ry, respectively to the same energy value at $\delta = 0$. For $20$ and $30$ GPa, new energy minima with $\Delta E = 7$ and $39$ \textit{meV}, observed at $\delta$ $\approx \pm 0.02$ and $\pm 0.03$ respectively. (b) TiO$_2$ structure with applied C$_{11}$-C$_{12}$ strain with axis $a=a_0(1 + \delta )$ and $b=a_0(1 - \delta )$.} 
\label{fig:tio2r_strain}
\end{figure}

The elastic combination $C_{11} - C_{12}$ corresponds to an orthorhombic shear distortion, which can be introduced through a symmetry-lowering strain. To linear order, this distortion transforms the lattice parameters as $a' = a_0(1+\delta)$, $b' = a_0(1-\delta)$, and $c' = c_0$, where $a_0$ and $c_0$ are the equilibrium tetragonal lattice constants and $\delta$ is the orthorhombic strain. The total energy of rutile TiO$_2$ as a function of $\delta$ at selected pressures is shown in Fig.~\ref{fig:strain_energy}.

At zero pressure, the energy exhibits a symmetric quadratic dependence on strain, with a minimum at $\delta = 0$, consistent with mechanical stability of the tetragonal phase. At 10~GPa, the curvature of the energy parabola is reduced, indicating a softening of the associated shear mode. At 20~GPa, the energy profile changes qualitatively: the high-symmetry point at $\delta = 0$ becomes a local maximum, and two symmetry-breaking minima emerge at approximately $\delta \approx \pm 0.02$. This behavior signals the onset of a spontaneous instability of the rutile phase toward an orthorhombic distortion. The energy lowering associated with the distorted configurations is about 7~meV per formula unit. At 30~GPa, the instability becomes more pronounced, with the minima shifting to $\delta \approx \pm 0.03$ and the stabilization energy increasing to $\sim$39~meV.

This behavior is characteristic of a tetragonal-to-orthorhombic shear instability and is consistent with the symmetry lowering associated with the CaCl$_2$-type structure \parencite{baur2007rutile}. The distorted structure at 30~GPa (Fig.~\ref{fig:tio2_pnnm}) exhibits the characteristic pattern of octahedral rotations, in which neighboring TiO$_6$ octahedra rotate in opposite directions. These rotations arise from displacements of oxygen atoms away from the high-symmetry $4f\,(u,u,0)$ positions toward $4g\,(x,y,0)$ sites with $x \neq y$ (Supplementary Materials). The combined effect of this internal distortion and the orthorhombic strain breaks the fourfold screw-axis symmetry of the $P4_2/mnm$ structure, lowering the symmetry to the orthorhombic $Pnnm$ phase. This corresponds to the same ferroelastic transition observed in other rutile-type systems.

The emergence of a double-well energy landscape as a function of the orthorhombic strain $\delta = \varepsilon_1 - \varepsilon_2$ (Fig.~2(a)) is characteristic of a continuous (second-order) phase transition and can be naturally described within a Landau framework \parencite{landau1937theory, carpenter1998elastic}. In this approach, the free energy is expanded in terms of a primary order parameter $Q$ as
\begin{equation}
E = \frac{1}{2} a (P - P_c) Q^2 + \frac{1}{4} b Q^4,
\end{equation}
where $a$ and $b$ are Landau coefficients, $P$ is the applied pressure, and $P_c$ is the critical pressure.

Identifying the orthorhombic strain $\delta$ as the primary order parameter, minimization of the free energy leads to the pressure dependence
\begin{equation}
\delta^2 =
\begin{cases}
0, & P < P_c, \\
\frac{a}{b} (P - P_c), & P > P_c.
\end{cases}
\end{equation}

\begin{figure}[htbp]
\centering
\includegraphics[width=0.8\linewidth]{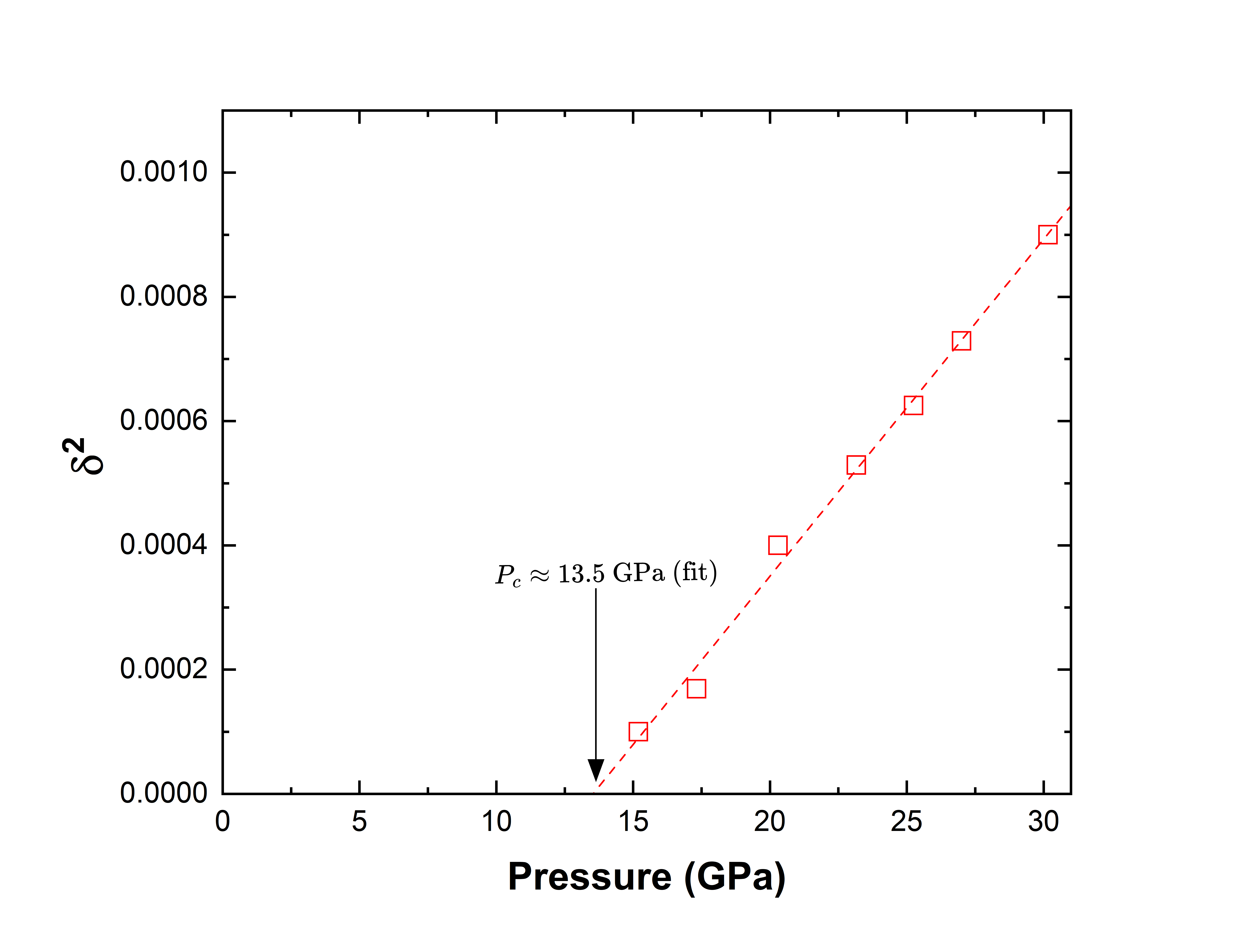}
\caption{Pressure dependence of the squared spontaneous strain $\delta^2$ in TiO$_2$. The dashed line represents a linear fit, highlighting the linear scaling $\delta^2 \propto (P - P_c)$ expected for a second-order transition.}
\label{fig:ess_pressure}
\end{figure}

The calculated pressure dependence of $\delta^2$ is shown in Fig.~\ref{fig:ess_pressure}. A clear linear relationship is observed for $P > P_c$, consistent with the Landau prediction for a second-order transition. Extrapolation of the linear fit yields a critical pressure of approximately 13.5~GPa for the onset of the $P4_2/mnm \rightarrow Pnnm$ transition. This behavior confirms that the orthorhombic strain acts as a proper order parameter for the transition, in agreement with the Landau prediction for a second-order transition.

This value is in good agreement with the instability pressure estimated from the generalized Born criterion, $C_{11} - C_{12} - 2P = 0$. The small discrepancy between the two estimates can be attributed to the harmonic approximation inherent in the elastic stability criterion, which is based on a second-order expansion of the elastic energy. In contrast, the strain–energy analysis explicitly captures anharmonic contributions to the free energy that become increasingly important near the phase transition, leading to a slightly lower critical pressure.

The elastic instability identified above can be naturally connected to lattice dynamics within the framework developed by Cowley \parencite{Cowley1976}. In structural phase transitions where a homogeneous deformation serves as the primary order parameter, the relevant fluctuations correspond to long-wavelength acoustic phonons. In this picture, the stability of the crystal is governed by the eigenvalues of the elastic constant matrix, while the associated fluctuations are embodied in acoustic modes whose velocities depend on these elastic constants. As an eigenvalue of the elastic stiffness tensor approaches zero, the velocity of the corresponding acoustic branch softens, signaling the onset of a lattice instability \parencite{Cowley1976}. This framework provides the basis for linking the pressure-induced elastic instability discussed above to the dynamical behavior of the phonon spectrum.

\begin{figure}[htbp]
\centering
\includegraphics[width=0.8\linewidth]{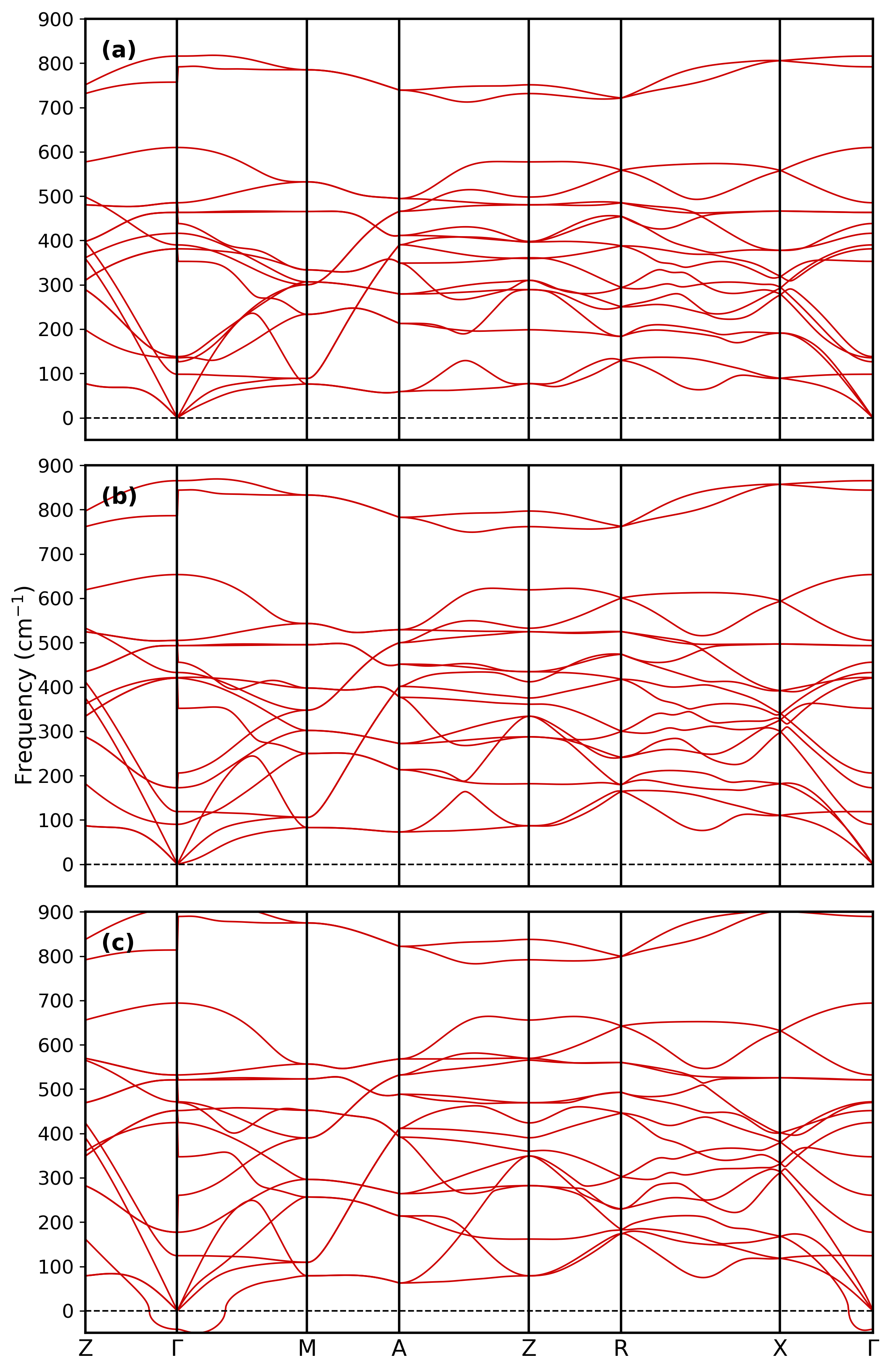}
\caption{Phonon dispersion relations of Titanium dioxide in the rutile phase calculated at different applied pressures: (a) 0 GPa, (b) 10 GPa, and (c) 20 GPa. Frequencies (cm$^{-1}$) are plotted along high-symmetry directions in the Brillouin zone (Z–$\Gamma$–M–A–Z–R–X–$\Gamma$).}
\label{fig:phonon}
\end{figure}

The evolution of the phonon dispersion of rutile TiO$_2$ under hydrostatic pressure provides a direct dynamical counterpart to the elastic instability discussed above. At ambient pressure (Fig.~\ref{fig:phonon}a), the phonon spectrum is fully stable, with all frequencies remaining positive throughout the Brillouin zone. The overall dispersion agrees well with neutron scattering measurements \parencite{Traylor1971}, confirming the reliability of the present calculations.

A quantitative comparison of $\Gamma$-point frequencies (Table~II) shows good agreement with experiment, particularly for Raman-active modes. For example, the calculated $B_{1g}$ frequency (138~cm$^{-1}$) is in closer agreement with experiment than previous calculations \parencite{changyol_lee_990c6044}. This establishes a reliable baseline for analyzing the pressure evolution of the soft mode associated with the structural instability.

Small discrepancies remain for the infrared-active transverse optical modes, particularly the low-frequency $E_u$ and $A_{2u}$ modes, which remain somewhat underestimated compared to experiment. Similar variations are found in previous first-principles studies, where calculated frequencies range from $\sim$144 \parencite{sikora2005ab} to $\sim$154~cm$^{-1}$ \parencite{montanari2002lattice}, depending on computational details. This spread reflects the strong sensitivity of polar modes to long-range electrostatic interactions, dielectric screening, and the equilibrium volume predicted by the exchange--correlation functional, with larger volumes leading to softer transverse optical modes. In addition, the harmonic approximation may contribute to the underestimation, as these low-frequency modes exhibit appreciable anharmonicity and temperature dependence. Despite these quantitative differences, the overall phonon spectrum and mode ordering are well reproduced.

The strong longitudinal--transverse optical (LO--TO) splitting at the $\Gamma$ point—most pronounced for the $A_{2u}$ and $E_u$ modes—is clearly reproduced. This splitting arises from long-range dipole--dipole interactions associated with highly polarizable Ti--O bonds \parencite{changyol_lee_990c6044}. The calculated Born effective charges further support this picture, with $Z^*_{Ti,xx} = 6.34$ and $Z^*_{Ti,zz} = 7.66$ for Ti, and $Z^*_{O,xx} = -3.17$ and $Z^*_{O,zz} = -3.83$ for O, indicating substantial dynamical charge transfer and pronounced anisotropy. Consistently, the high-frequency dielectric tensor yields $\epsilon_{\infty,xx} = 7.52$ and $\epsilon_{\infty,zz} = 8.90$, confirming enhanced dielectric response along the $c$-axis. The inequalities $Z^*_{Ti,zz} > Z^*_{Ti,xx}$ and $\epsilon_{\infty,zz} > \epsilon_{\infty,xx}$ highlight the intrinsic anisotropy of the rutile lattice and validate the treatment of long-range electrostatics in the present calculations.

\begin{table}[h]
\centering
\caption{Calculated and experimental $\Gamma$ point vibrational modes of rutile TiO$_2$ at ambient pressure. All values in cm$^{-1}$.}
\begin{tabular}{lccc}
\hline
\textbf{Mode} & \textbf{This work} & \textbf{Ref.\,\parencite{changyol_lee_990c6044}} & \textbf{Experiment} \\
\hline
\multicolumn{4}{l}{\textbf{Raman}} \\
$B_{1g}$      & 138 & 125 & 142$^{a}$, 143$^{b}$ \\
$E_g$         & 463 & 472 & 445$^{a}$, 447$^{b}$ \\
$A_{1g}$      & 609 & 623 & 610$^{a}$, 612$^{b}$ \\
$B_{2g}$      & 815 & 828 & 825$^{a}$, 827$^{b}$ \\
\hline
\multicolumn{4}{l}{\textbf{Infrared}} \\
$E_u$ (TO)    & 146 & 165 & 189$^{a}$, 183$^{c}$ \\
$E_u$ (LO)    & 349 & 352 & 375$^{a}$, 373$^{c}$ \\
$E_u$ (TO)    & 374 & 391 &  388$^{c}$ \\
$E_u$ (LO)    & 438 & 442 & 429$^{a}$, 458$^{c}$ \\
$E_u$ (TO)    & 491 & 493 & 494$^{a}$, 500$^{c}$ \\
$E_u$ (LO)    & 792 & 808 & 842$^{a}$, 807$^{c}$ \\
$A_{2u}$ (TO) & 136 & 176 & 173$^{a}$, 167$^{c}$ \\
$A_{2u}$ (LO) & 760 & 769 &  812$^{c}$ \\
\hline
\multicolumn{4}{l}{\textbf{Silent}} \\
$A_{2g}$      & 416 & 416 & - \\
$B_{1u}$      & 116 & 117 & 113$^{a}$ \\
$B_{1u}$      & 391 & 408 & 406$^{a}$ \\
\hline
\end{tabular}

\vspace{0.5em}

\noindent{\footnotesize{
$^{a}$ Ref.~\parencite{Traylor1971}; 
$^{b}$ Ref.~\parencite{Porto1967}; 
$^{c}$ Ref.~\parencite{Eagles1964}
}}

\end{table}

Upon compression to 10~GPa (Fig.~\ref{fig:phonon}b), most phonon branches exhibit the expected hardening due to increasing interatomic force constants (Supplementary Materials). In contrast, the low-frequency $B_{1g}$ mode softens significantly, decreasing from 138 to 90~cm$^{-1}$. This behavior is consistent with experimental observations of anomalous pressure dependence \parencite{nicol1971raman, samara1973pressure} and signals the onset of a structural instability associated with this mode.

At 20~GPa (Fig.~\ref{fig:phonon}c), the instability becomes explicit: the $B_{1g}$ branch develops imaginary frequencies at the $\Gamma$ point and along symmetry directions. The appearance of imaginary frequencies indicates that the rutile structure is dynamically unstable and will spontaneously distort to a lower-symmetry configuration. This provides direct dynamical confirmation of the instability inferred from the elastic constants.

The softening of the $B_{1g}$ mode is consistent with the elastic instability governed by the $C_{11} - C_{12}$ combination. As this elastic modulus approaches zero, the resistance to orthorhombic shear vanishes, and the corresponding lattice vibration becomes unstable. In this sense, the $B_{1g}$ mode represents the dynamical manifestation of the same instability that drives the ferroelastic distortion. The associated displacement pattern involves coordinated in-plane motion of oxygen atoms, producing alternating rotations of TiO$_6$ octahedra about the $c$-axis. This distortion preserves inversion symmetry but breaks the fourfold rotational symmetry, consistent with the transformation from $P4_2/mnm$ to $Pnnm$ \parencite{montanari2004pressure}.

Our results help clarify discrepancies with previous theoretical studies. Lukačević \textit{et al.} \parencite{igor_luka_evi__5df20077} reported a softening of the $B_{1g}$ mode and a negative Grüneisen parameter, and interpreted the resulting instability as a transition toward the columbite ($Pbcn$) structure at relatively low pressures (5--10~GPa). In contrast, our calculations show that the rutile phase remains dynamically stable up to significantly higher pressures, consistent with the persistence of real phonon frequencies up to $\sim$15~GPa. Moreover, the symmetry of the unstable mode corresponds to an orthorhombic distortion leading to the $Pnnm$ phase, rather than the columbite structure. This difference likely reflects the sensitivity of the structural assignment to the symmetry of the soft phonon mode and its coupling to lattice strain.

Taken together, these results provide a unified picture of the transition mechanism. The pressure-induced softening of the $B_{1g}$ phonon mode, the collapse of the $C_{11} - C_{12}$ elastic modulus, and the emergence of a double-well energy landscape with respect to orthorhombic strain all point to the same underlying instability. The rutile-to-$Pnnm$ transition is driven by the coupling between a zone-center lattice vibration and a homogeneous shear strain, with the phonon instability providing the dynamical pathway through which the lattice relaxes into the lower-symmetry configuration. This correspondence establishes the $B_{1g}$ mode as the primary lattice instability associated with the ferroelastic order parameter.

While the preceding elastic and phonon analyses identify the CaCl$_2$-type phase as the primary product of the instability, the experimentally observed phase is ultimately determined by thermodynamic stability. This is governed by the enthalpy, $H = E + PV$, relative to competing polymorphs. To evaluate this, the pressure–volume relationships were fitted using a third-order Birch–Murnaghan equation of state \parencite{birch1947finite}, enabling a direct comparison of enthalpies among the rutile, CaCl$_2$-type ($Pnnm$), columbite ($Pbcn$), and baddeleyite ($P2_1/c$) phases (Supplementary Materials).

\begin{figure}[htbp]
\centering
\includegraphics[width=0.8\linewidth]{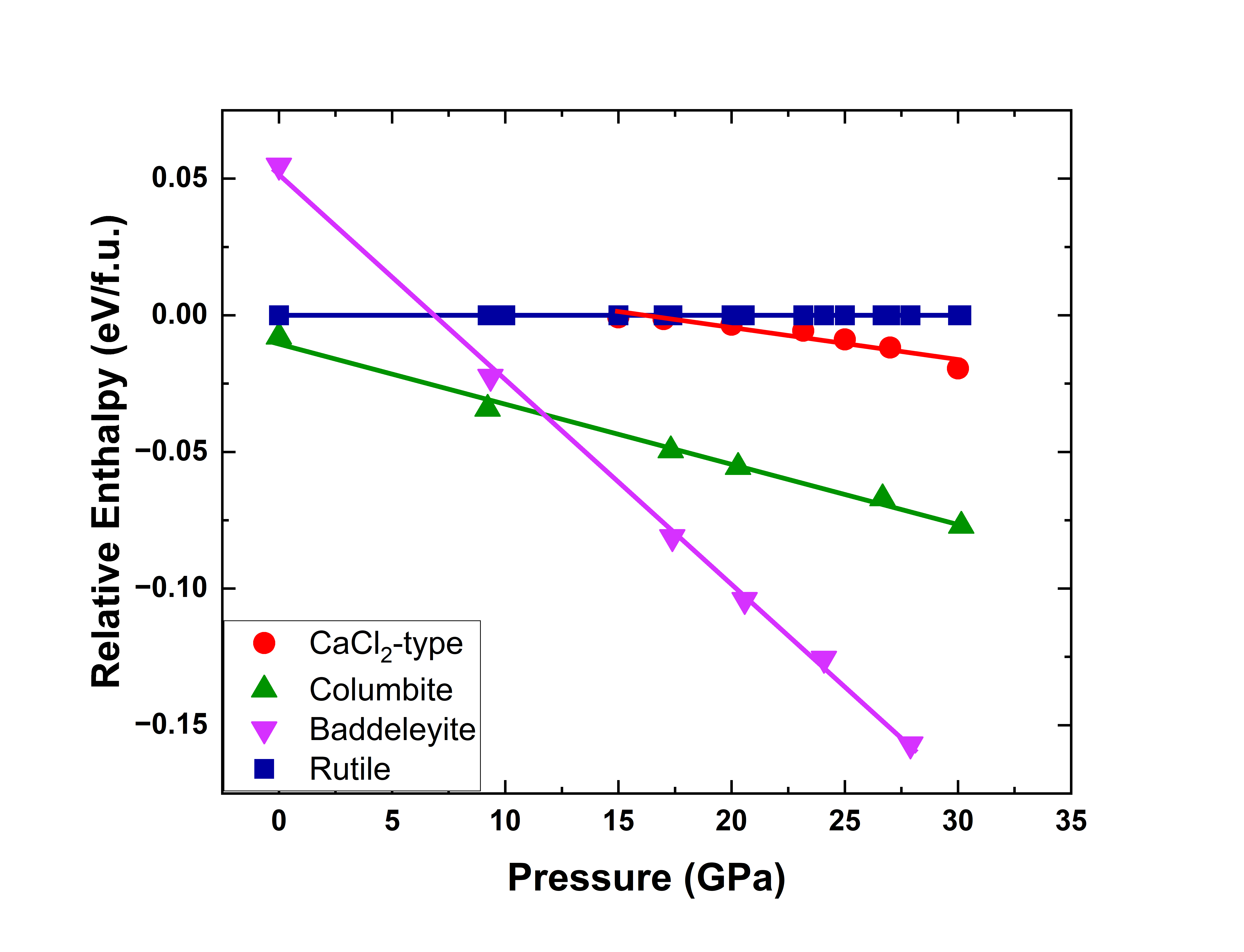}
\caption{Pressure dependence of the relative enthalpy per formula unit for different TiO$_2$ polymorphs, referenced to the rutile phase. Results are shown for rutile, Pnnm, columbite, and baddeleyite structures over the pressure range 0–30 GPa.}
\label{fig:enthalpy}
\end{figure}

The pressure dependence of the relative enthalpies is shown in Fig.~\ref{fig:enthalpy}. At low pressures, the columbite phase is energetically favored relative to rutile, with a stabilization of up to $\sim$143~meV per formula unit below $\sim$11.7~GPa. Upon further compression, the baddeleyite phase becomes the most stable, and its enthalpy advantage increases steadily with pressure, consistent with previous results obtained using GGA \parencite{ma2009pressure}, hybrid functionals \parencite{swamy2014first}, and meta-GGA approaches \parencite{pohlmann2025stability}. 

In contrast, the CaCl$_2$-type phase—although stabilized by the ferroelastic instability identified above—remains only weakly stabilized relative to rutile. Even at 20~GPa, its stabilization energy is limited to $\sim$7~meV per formula unit, which is significantly smaller than that of the baddeleyite phase (on the order of $\sim$402~meV). This small thermodynamic gain for the CaCl$_2$-type structure arises almost entirely from the internal energy $E$, as its volume remains nearly identical to that of rutile. By contrast, the columbite and baddeleyite phases exhibit substantial volume reductions of approximately $4$ and $15$~bohr$^3$ per formula unit, respectively. These reductions lead to significant contributions to the $PV$ term—on the order of $0.5$ and $1.9$~eV—which more than compensate for the increase in $E$, thereby stabilizing these phases under pressure.

This behavior contrasts with that of other rutile-type dioxides. For example, in CrO$_2$, the CaCl$_2$-type phase is stabilized over a broad pressure range and becomes energetically competitive with rutile above $\sim$15~GPa \parencite{b__r__maddox_82081e3d}. In that system, the columbite phase does not emerge as the ground state until significantly higher pressures ($\sim$35.5~GPa), followed by a transition to the pyrite structure above $\sim$45~GPa \parencite{li2012structural, bendaoud2019predicted}. Similarly, in GeO$_2$, the CaCl$_2$-type phase is stabilized over an extended pressure window, becoming the dominant phase above $\sim$19~GPa and remaining stable up to $\sim$36~GPa before transitioning to the columbite phase \parencite{lodziana2001ab}. A similar stabilization of the CaCl$_2$-type phase has also been observed in $\beta$-MnO$_2$, where experimental studies report a rutile-to-$Pnnm$ transition at low pressures \parencite{haines1995second, curetti2019low}, and first-principles calculations predict that this phase persists over an extended pressure range before higher-coordination structures become favorable \parencite{li2006ab}.

The distinct behavior of TiO$_2$ can be traced to the relative ordering of competing high-pressure polymorphs. In contrast to CrO$_2$ and GeO$_2$, both the columbite and baddeleyite phases of TiO$_2$ are stabilized at comparatively low pressures. As a result, these phases suppress the CaCl$_2$-type structure, confining it to a metastable region of the phase diagram.

This competition provides a natural explanation for the long-standing absence of a clearly resolved $Pnnm$ phase in experiments. Although the CaCl$_2$-type structure emerges as the immediate product of the ferroelastic instability, it is rapidly overtaken by more stable high-coordination phases under typical experimental conditions. Consequently, the CaCl$_2$-type phase in TiO$_2$ should be regarded not as a stable equilibrium phase, but rather as a transient or weakly metastable intermediate along the transformation pathway. 


\section{Conclusion}


In this work, we have presented a unified description of the pressure-induced instability of rutile TiO$_2$ by combining elastic, lattice dynamical, and thermodynamic analyses. Our results show that the instability of the rutile phase is driven by a ferroelastic shear mode associated with the $C_{11} - C_{12}$ elastic combination. This macroscopic instability is accompanied by the softening and eventual condensation of the $B_{1g}$ phonon mode, whose displacement pattern corresponds to coordinated rotations of TiO$_6$ octahedra. The emergence of a double-well energy landscape with respect to orthorhombic strain further confirms that the transition can be described within a Landau framework, with the strain acting as the primary order parameter.

Taken together, these results establish that the CaCl$_2$-type ($Pnnm$) structure is the natural symmetry-lowered product of the instability. In this sense, the $Pnnm$ phase represents the intrinsic outcome dictated by both elasticity and lattice dynamics. However, our enthalpy analysis reveals that this phase is only weakly stabilized relative to rutile and remains significantly less favorable than competing high-pressure polymorphs, particularly the columbite and baddeleyite phases. As a result, the $Pnnm$ phase is thermodynamically suppressed over the relevant pressure range.

This competition provides a consistent explanation for the absence of a clearly resolved $Pnnm$ phase in experiments. Although the CaCl$_2$-type structure emerges as the primary product of the ferroelastic instability, it does not establish itself as a stable equilibrium phase under compression. Instead, it should be regarded as a weakly metastable or transient intermediate that is readily bypassed as the system evolves toward more stable, higher-coordination structures.

More broadly, this work highlights the importance of considering both dynamical instabilities and thermodynamic competition in understanding pressure-induced phase transitions. The framework developed here may be extended to other rutile-type dioxides, where a similar interplay between symmetry, lattice dynamics, and enthalpy governs whether intermediate phases emerge as stable polymorphs or remain experimentally elusive.

%
%

\ack{J.P. and A.-J.O. are thankful for travel support from the Department of Physics and the School of Graduate Studies at Baylor University. The authors thank C. Bell and M. Hutcheson at the High Performance Computing Center of Baylor University for technical support.}

\funding{None.}

\roles{
J.P.: Data curation, Investigation, Formal analysis, Validation, Methodology, Writing – original draft.\newline
A.-J.O.: Data curation, Investigation, Formal analysis, Validation.\newline
K.P.: Conceptualization, Investigation, Writing – review and editing, Supervision.
}

\data{The data supporting the findings of this study are available in the Supplementary Materials.}

\suppdata{Additional data including equilibrium structures, fit to Birch–Murnaghan equation of states, structural parameters of the CaCl$_2$-type phase at selected pressures, and high-pressure phonon frequencies are provided in the Supplementary Materials.}

\printbibliography

\end{document}